\documentstyle[sprocl,epsfig]{article}

\def\be{\begin{equation}}
\def\ee{\end{equation}}
\def\bea{\begin{eqnarray}}
\def\eea{\end{eqnarray}}
\def\epem{e^+e^-}

\begin{document}

\title{NOVEL 3D CLUSTERING ALGORITHM AND 
TWO PARTICLE SEPARATION WITH TILE HCAL}

\author{V.MORGUNOV$^1$, A.RASPEREZA$^2$ }

\address{$^1$ITEP, B.Cheremushkinskaja 25, 117218 Moscow Russia \\
$^2$DESY-Hamburg, Notkestr. 85, D-22607 Hamburg Germany}

\maketitle

\begin{center}
{\it{Proceedings of the International Conference on Linear Colliders}}\\
{\it{LCWS 2004, Paris, 19$-$23 April 2004}}
\end{center}

\vspace{5mm}

\abstracts{
Based on a novel shower reconstruction algorithm,  
the study of two particle separation with tile hadron
calorimeter (HCAL) is performed. 
The separability of two close-by particles is 
found to be strongly dependent on transverse and
longitudinal segmentation of tile HCAL. 
}

\section{Introduction}
Event reconstruction in a future linear
$\epem$ collider experiment will be based on a particle 
flow concept~\cite{pflow} aiming at
reconstruction of every single particle in an event. 
Clearly, performance of the particle flow algorithm will  
strongly depend on its capability to separate two and more close-by 
showers. In its turn particle separation capability will be influenced 
by transverse and longitudinal segmentation of calorimeters.
In this note we investigate the impact of transverse and 
longitudinal segmentation of hadron calorimeter on two particle separation.
The study is based on a novel shower reconstruction algorithm which 
makes use of finely granulated electromagnetic and hadron calorimeters
foreseen for a linear collider detector.

\section{Simulation}
The detector setup used in our simulation closely follows 
the calorimeter design outlined
in the TESLA Technical Design report~\cite{TDR}. 
The elecromagnetic calorimeter consists of tungsten absorbers interspersed 
with silicon diod pads and is characterised by a very high granularity.
The transverse size of the readout cell is 1$\times$1 cm$^2$.
The hadron calorimeter represents an analog device, consisting of 
stainless steel absorber plates interspersed with scintillating 
tiles. Calorimeter parameters are given in Table~\ref{tab:calo}.
\begin{table}[ht]
\begin{center}
\begin{tabular}{|c|c|c|c|c|c|}
\hline
Calorimeter & Type    & Number of & Thickness of         & Thickness of \\
            &         & layers    & absorber layers      & active layer \\
\hline
ECAL        & W/Si    & 40        & $\phantom{0}$1-30 : 1.4mm  & 0.5mm  \\
            &         &           & 31-40 : 4.2mm        &              \\
\hline
HCAL        & Fe/Sci  & 40        & 20mm                 & 5mm          \\
\hline
\end{tabular}
\end{center}
\caption{Parameters of calorimeters.}
\label{tab:calo}
\end{table}
In our studies calorimeter response has been simulated
using the GEANT3 package~\cite{GEANT3}. Hadronic interactions 
are simulated using FLUKA~\cite{FLUKA} complemented
with low energy neutron transport code MICAP~\cite{MICAP}. 

Several options of transverse granularity of HCAL have been
considered: 1$\times$1, 3$\times$3 and 5$\times$5 cm$^2$. 
For the 3$\times$3 cm$^2$ tile size, the readout scheme with 
each two adjacent layers joined in depth is also considered.  
The influence of transverse tile size is studied using a
recently developed shower reconstruction procedure which takes
advantage of the fine granularity of calorimeters.

\section{Clustering}
Before describing clustering and the shower reconstruction procedure, 
some definitions need to be introduced. 
By clusters we mean internal structures inside 
shower, e.g. pieces of tracks produced
by minimal ionising particles, electromagnetic 
subshowers originated from neutral pions or set
of adjacent hits produced by several charged particles
in the vicinity of nuclear interaction points.
The shower is then viewed as a group of topologically 
connected clusters.

\begin{figure}[t]
\begin{minipage}[c]{0.48\textwidth}
\includegraphics*[width=0.99\textwidth]{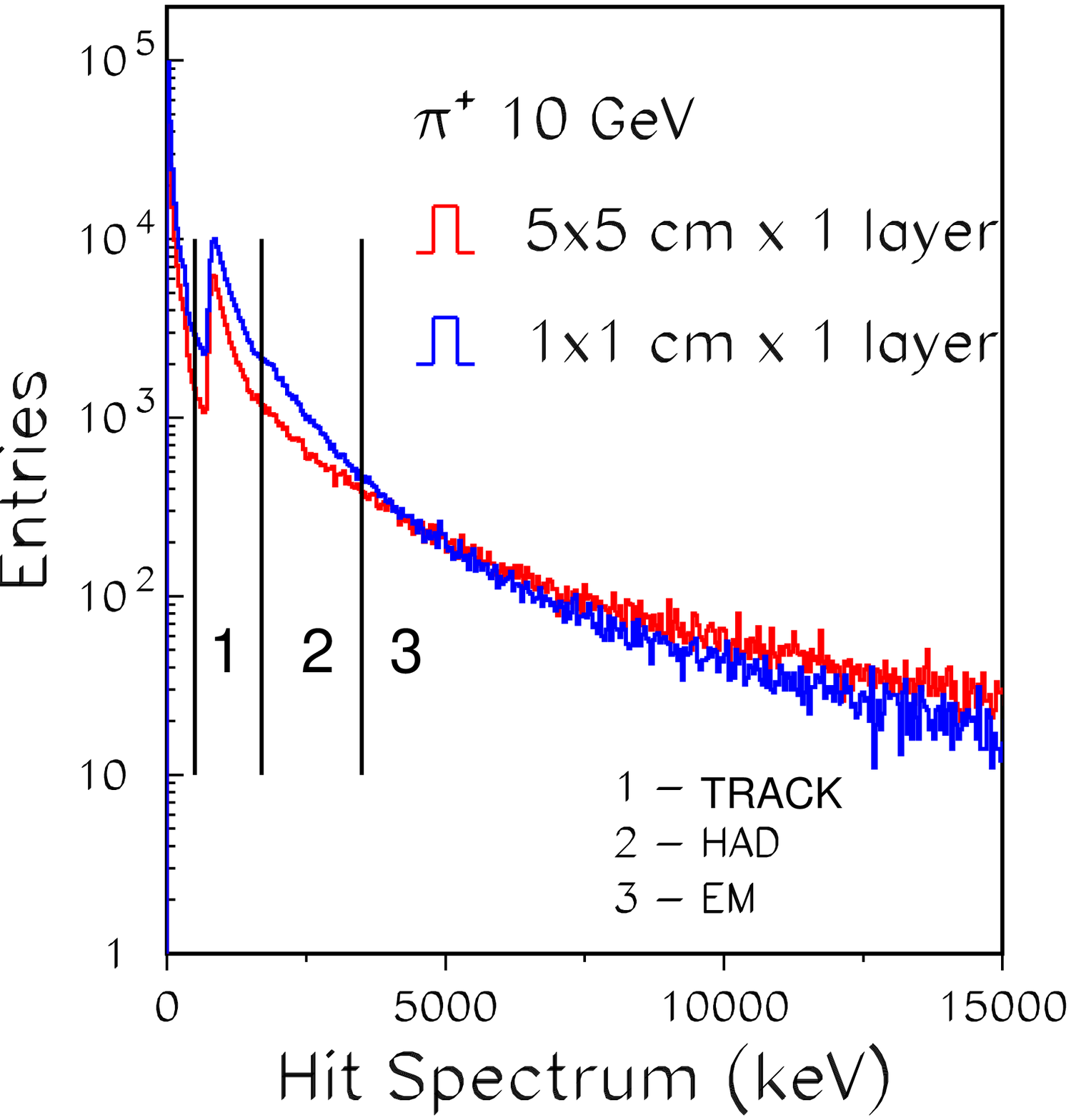}
\caption{Hit classification based on 
hit energy.
\label{fig:hits}}
\end{minipage}
\begin{minipage}[c]{0.02\textwidth}
.
\end{minipage}
\begin{minipage}[c]{0.48\textwidth}
\includegraphics*[width=0.99\textwidth]{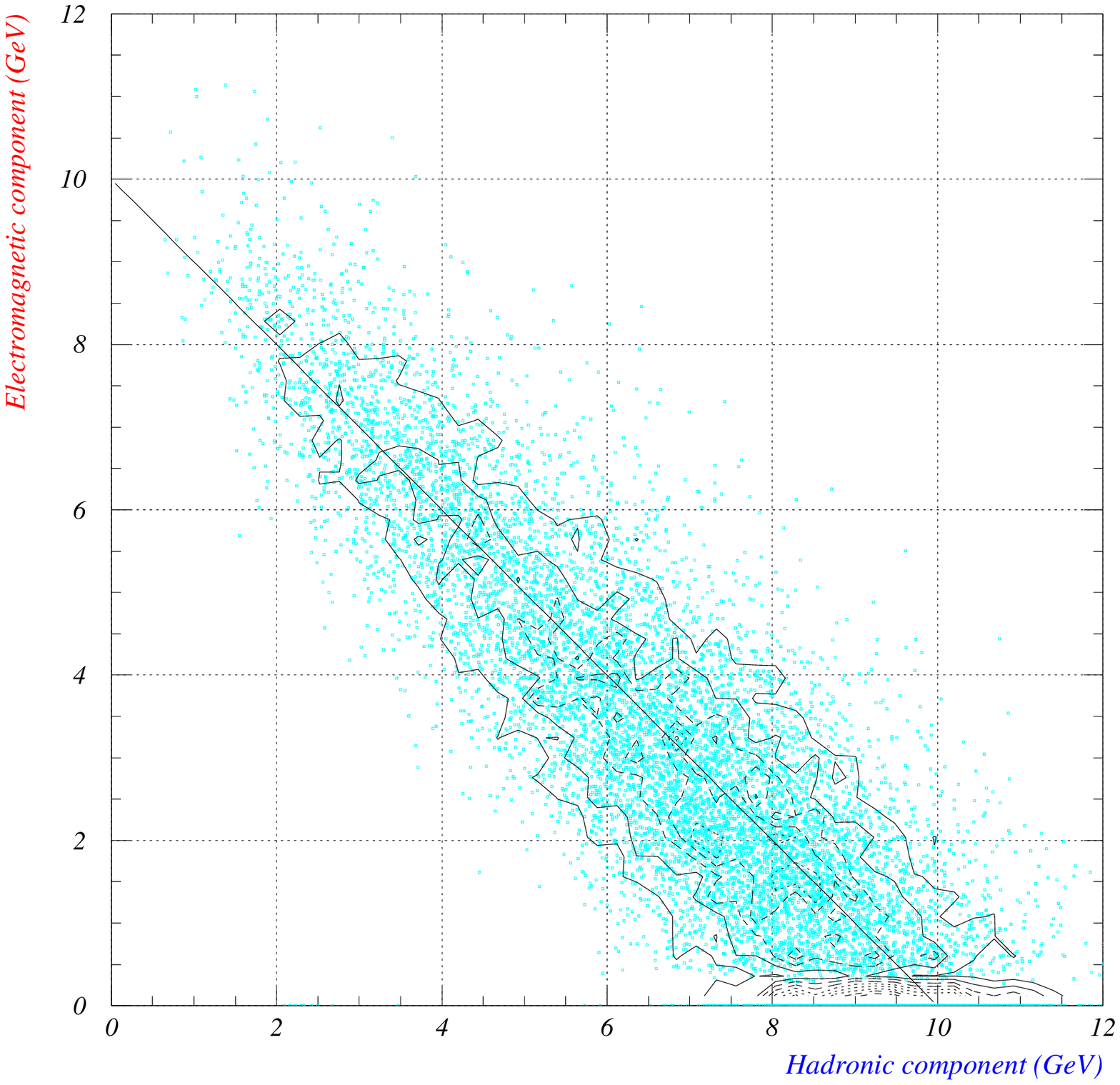}
\caption{Correlation between electromagnetic and hadronic
components of a 10 GeV $K^0_S$ shower.
\label{fig:had_em}}
\end{minipage}
\end{figure}

Clustering begins with the hit 
classification procedure based on the energy  
of each hit. Hits with energy deposit greater 
than half a MIP signal and less than 1.7 MIPs are 
assigned for the so called "track--like" category. 
Hits with an extremely dense energy deposit exceeding
3.5 times MIP expectation, are considered as 
relics of electromagnetic activity. Finally, 
hits with energy deposit ranging from 1.7  
to 3.5 MIPs are assigned for "hadron--like" category. 
Hit classification is illustrated in Figure~\ref{fig:hits}.
At the next stage, the clustering procedure based
on 3D pattern recognition is performed and clusters are
classified into different categories taking into account
hit categorisation, topological properties
of clusters, their inter--relations and position in 
the calorimeter volume. The "track--like" clusters are 
classified as having small eccentricity and low hit density.
The "hadron--like" clusters have relatively large
eccentricity and low hit density. The "electromagnetic--like" 
clusters have high hit density and small eccentricity.
An additional hit category is introduced 
by clustering. These are hits spatially
disconnected from other clusters and presumably 
initiated by neutrons, hence the name for 
this category: "neutron--like" hits. 

\begin{figure}[t]
\begin{minipage}[c]{0.46\textwidth}
\vspace*{-15mm}
\includegraphics*[width=0.99\textwidth]{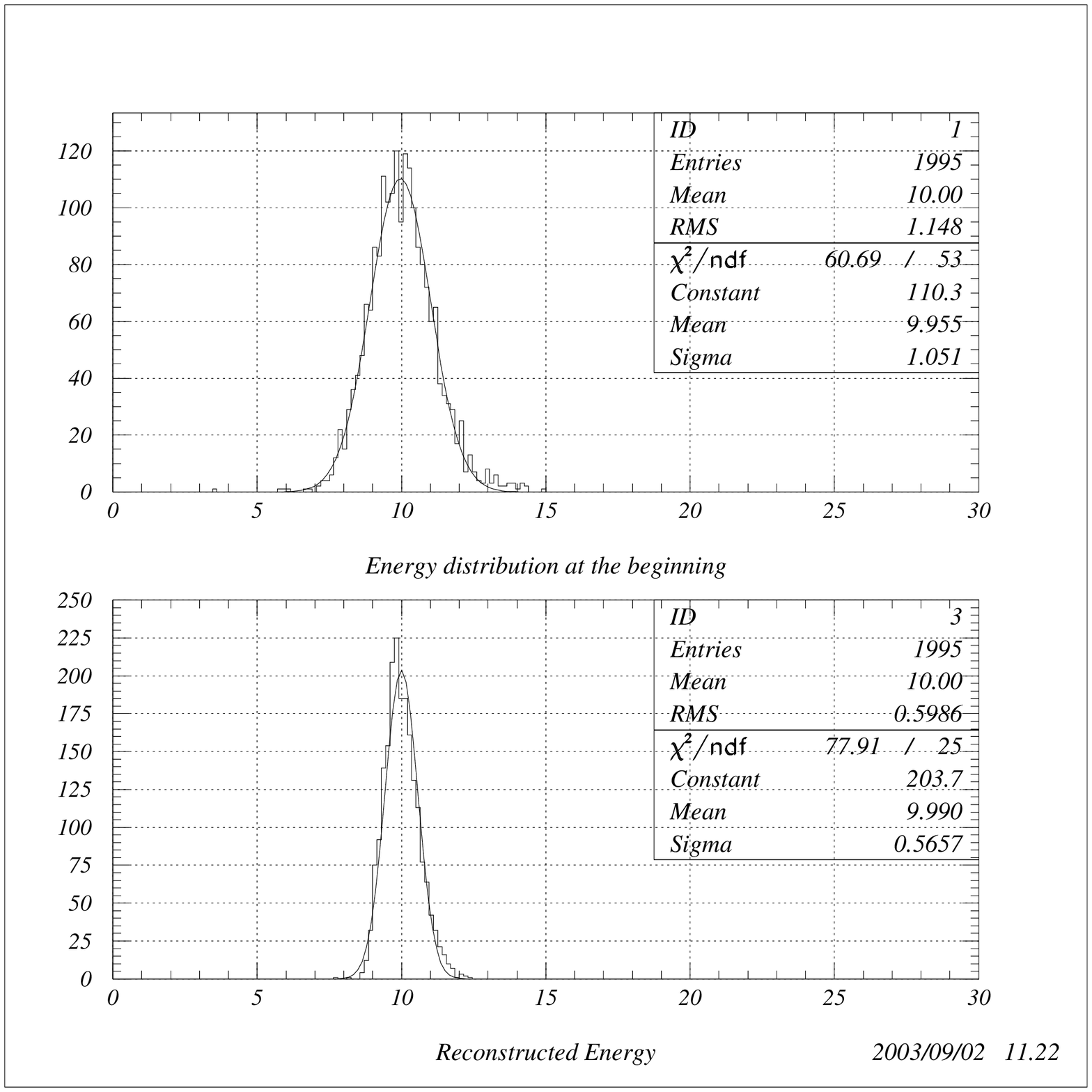}
\vspace*{-15mm}
\caption{Hadron energy resolution without (upper plot) and with 
(lower plot) weighting procedure using a priori knowledge of 
hadron beam energy.
\label{fig:ecorr}}
\end{minipage}
\begin{minipage}[c]{0.04\textwidth}
.
\end{minipage}
\begin{minipage}[c]{0.42\textwidth}
\begin{center}
\includegraphics*[width=0.99\textwidth]{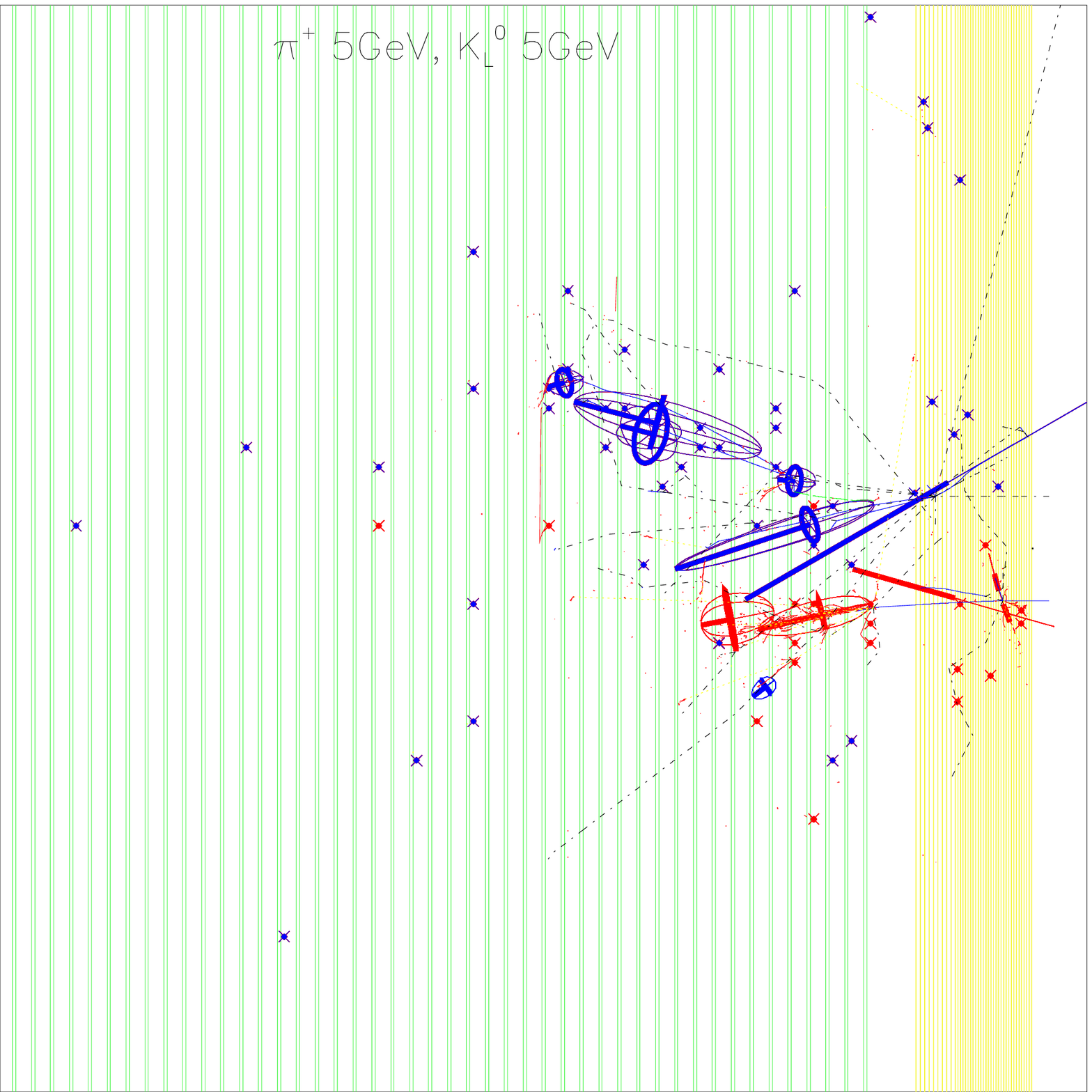}
\caption{Two reconstructed showers from 5 GeV $K^0_L$ and 5 GeV $\pi^+$.
Showers are composed of clusters represented by inertia ellipsoids.
Neutron hits are indicated as a cross-dots.
\label{fig:showers}}
\end{center}
\end{minipage}
\end{figure}

Such a clustering procedure results in separation of total
energy into different components. Correlation between hadronic component,
including energy contained in the "track--like", "hadron--like" clusters
and "neutron--like" hits and electromagnetic component, including
energy contained in the "electromagnetic--like" showers, is illustrated 
in Figure~\ref{fig:had_em}. 
The reconstructed energy distribution 
of a 10 GeV $\pi^+$ shower after dedicated energy correction procedure
similar to that used by DREAM collaboration~\cite{DREAM} is presented in
Figure~\ref{fig:ecorr}. This energy correction procedure 
is based on individual weighting of hadronic and electromagnetic
components of the shower energy and uses a priori knowledge of 
the $\pi^+$ beam energy. Such an approach is obviously non-Bayesian
and Figure~\ref{fig:ecorr} indicates only the degree of correlation 
between measured hadronic and electromagnetic components
of the shower energy rather than 
realistically achievable energy resolution.

\section{Shower Reconstruction}
Once clustering is performed, showers are reconstructed  
by building a "tree" of "electromagnetic--like" and "hadron--like" clusters 
connected by "track--like" clusters. "Neutron--like" hits and clusters 
disconnected from the main tree are added to shower if 
their distance to the shower axis is less than some cut parameter,
$D_{cut}$. Shower axis is defined at 
each step of the shower building process as the main principle axis 
of inertia tensor associated with shower. 
For showers produced by charged particles,
the cluster nearest to the track intersection point with the ECAL
front plane seeds the shower. Furthermore, for showers initiated
by charged particles, parameter $D_{cut}$ is adjusted iteratively 
during the process of shower building until
the energy contained in the 
reconstructed shower matches the best the momentum of the associated track.
Typical value of $D_{cut}$ ranges from 1 to 3 cm.

\section{Two Particle Separation}
\begin{figure}[t]
\begin{minipage}[c]{0.45\textwidth}
\includegraphics*[width=0.99\textwidth]{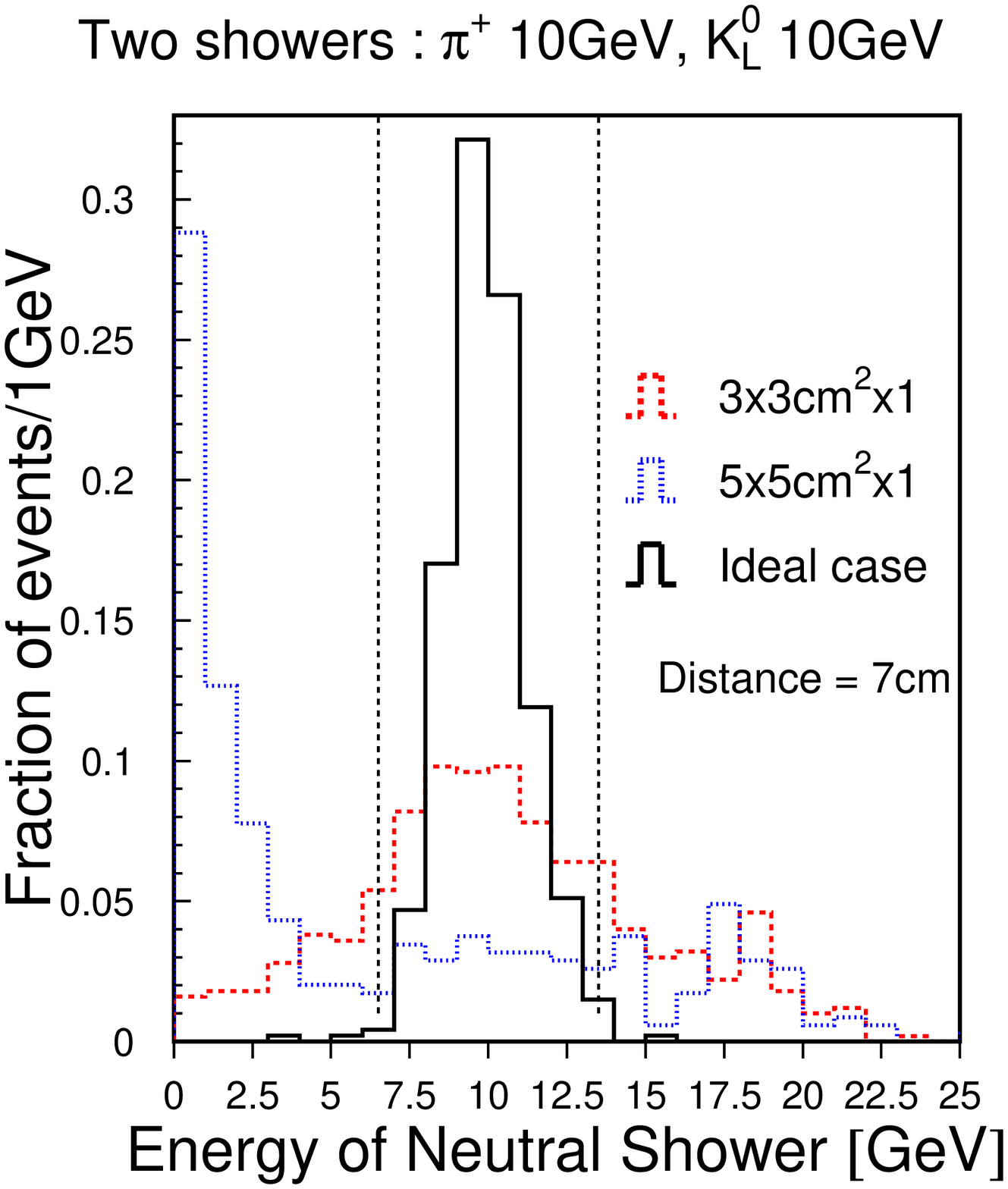}
\caption{Distributions of the reconstructed energy of the
neutral shower from $K^0_S$ in the presence of nearby shower from 
$\pi^+$ for two options of HCAL transverse segmentation,  
3$\times$3 and 5$\times$5 cm$^2$ (dashed and dotted histograms, respectively).
Solid histogram shows reference distribution obtained for a 
10 GeV $K^0_S$ shower in the absence of nearby shower. 
\label{fig:energy}}
\end{minipage}
\begin{minipage}[c]{0.02\textwidth}
.
\end{minipage}
\begin{minipage}[c]{0.45\textwidth}
\includegraphics*[width=0.99\textwidth]{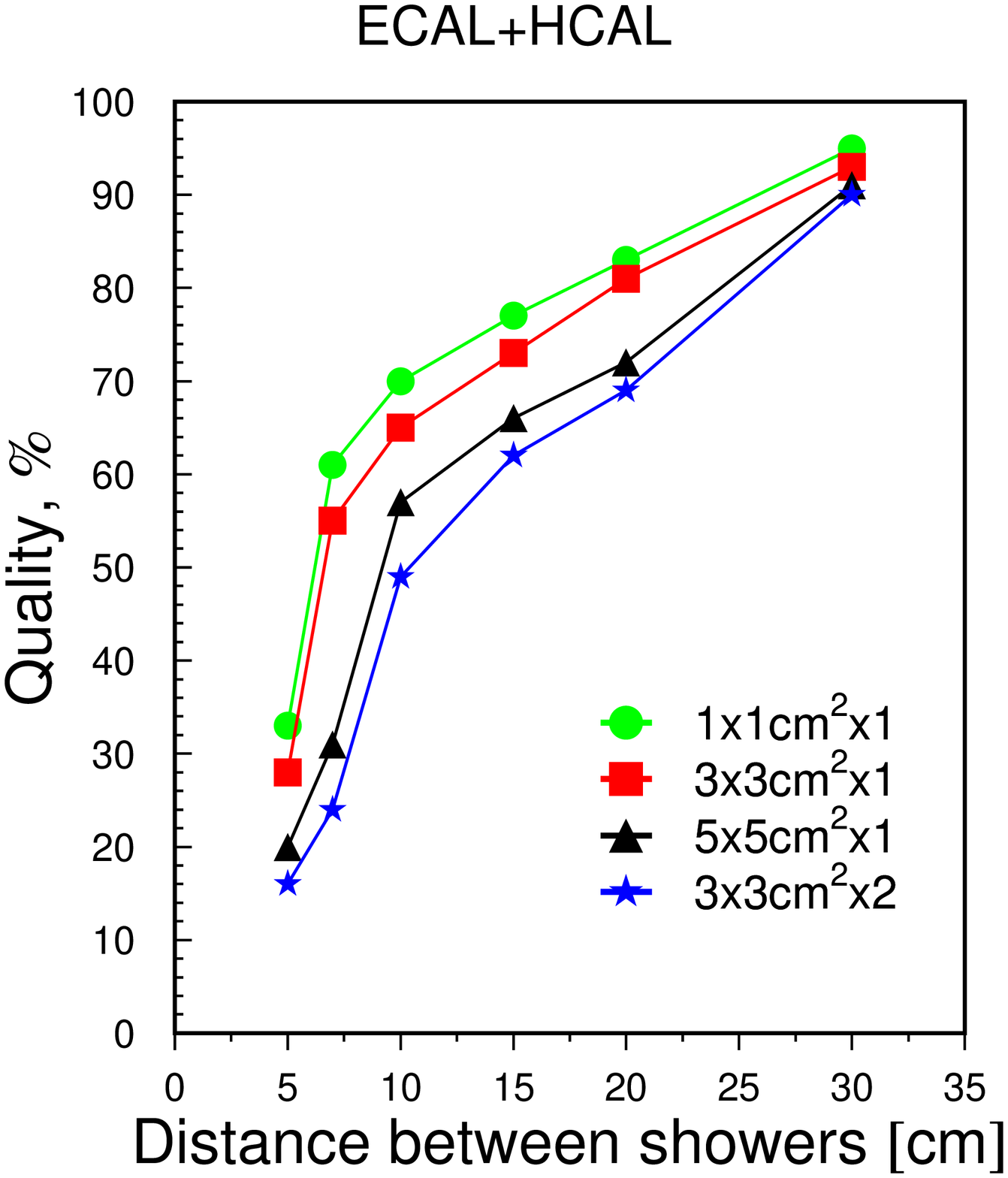}
\caption{Separation quality as a function of distance
between two particles, 10 GeV $K^0_S$ and 10 GeV $\pi^+$,  
for different options of HCAL transverse and longitudinal 
segmentation.
\label{fig:quality}}
\end{minipage}

\end{figure}

To estimate shower separation performance we simulated
the response of the electromagnetic and hadron calorimeters to two particles, 
$K^0_S$ and $\pi^+$, which are normally incident at the ECAL front plane. 
Our preliminary study did not take into account the effect of a magnetic 
field. The momentum of $\pi^+$ and its trajectory are 
assumed to be precisely measured with an inner tracking system. 
The particle separation performance is estimated
as a function of distance between two particles. 
Figure~\ref{fig:showers} presents an example of two resolved 
showers initiated by $K^0_S$ and $\pi^+$ with an energy of 5 GeV each.
Figure~\ref{fig:energy} shows the distribution of the reconstructed 
energy of the neutral shower produced by a 10 GeV $K^0_L$ in the 
proximity of the shower produced by a 10 GeV $\pi^+$. 
The distance between the two particles is 7cm. Distributions are shown for 
two options of transverse segmentation of an analog HCAL, 
3$\times$3 and 5$\times$5 cm$^2$.
The reference energy distribution obtained for 
a 10 GeV $K^0_L$ shower in the absence of any nearby shower is also shown. 
Performance is quantified in terms of particle separation quality defined
as a fraction of events in which the reconstructed energy of 
a neutral shower lies in the interval $E_{true}\pm 3\sigma$, 
where $E_{true}$ is the true energy of $K^0_L$ and $\sigma$
is the nominal energy resolution.
The separation quality is found to be hihgly sensitive to both 
transverse and longitudinal segmentation of the HCAL as demonstrated
in Figure~\ref{fig:quality}. An independent approach of shower
separation based on an alternative clustering method with 
minimal spanning trees
gives comparable results for a digital calorimeter 
with 1$\times$1 cm$^2$ RPC cell size~\cite{span_tree}.

\end{document}